\documentclass{llncs}
\usepackage{makeidx}  
\usepackage[linesnumbered]{algorithm2e}
\usepackage{graphicx}
\usepackage{amsmath}
\usepackage{multirow,array}
\usepackage{lstsemantic}
\usepackage{amssymb}
\PassOptionsToPackage{hyphens}{url}\usepackage{hyperref}

\usepackage[utf8]{inputenc}

\newcommand{\tttt}[1]{\texttt{\small #1}}

\newcommand{\nameit}[1]{\ensuremath{\mathit{#1}}\xspace}

\newcommand{\namecal}[1]{\ensuremath{\mathcal{#1}}\xspace}
\newcommand{\name}[1]{\nameit{#1}}

\usepackage{xifthen}

\newcommand{\concat}[2][]{%
  \ifthenelse{\isempty{#1}}%
    {\name{#2}}
    {\name{#2#1}}
}

\newcommand{\sub}[2][]{%
  \ifthenelse{\isempty{#1}}%
    {\name{#2}}
    {\name{#2_\name{#1}}}
}

\newcommand{\HDT}[1][]{\sub[#1]{HDT}}
\newcommand{\Sec}[1][]{\sub[#1]{Sec}}
\renewcommand{\H}[1][]{\sub[#1]{H}}
\newcommand{\D}[1][]{\sub[#1]{D}}
\newcommand{\T}[1][]{\sub[#1]{T}}
\newcommand{\SO}[1][]{\sub[#1]{SO}}
\renewcommand{\S}[1][]{\sub[#1]{S}}
\renewcommand{\P}[1][]{\sub[#1]{P}}
\renewcommand{\O}[1][]{\sub[#1]{O}}
\newcommand{\s}[1][]{\sub[#1]{s}}
\newcommand{\p}[1][]{\sub[#1]{p}}
\renewcommand{\o}[1][]{\sub[#1]{o}}
\newcommand{\triple}{\name{(s,p,o)}}
\newcommand{\ID}[1][]{\sub[#1]{ID}}

\newcommand{\Seq}[1][]{\sub[#1]{Q}}
\newcommand{\Bit}[1][]{\sub[#1]{B}}

\newcommand{\terms}{\namecal{N}}
\newcommand{\I}{\namecal{I}}
\newcommand{\B}{\namecal{B}}
\renewcommand{\L}{\namecal{L}}
\newcommand{\G}{\name{G}}

\newcommand{\mapping}[2][]{%
  \ifthenelse{\isempty{#1}}%
    {M(#2)}
    {M(#2,#1)}
}

\newcommand{\cat}{cat}

\usepackage{todonotes} 

\definecolor{mygreen}{HTML}{00CC00}

\begin{document}

\title{HDTCat: let's make HDT scale}
\titlerunning{HDTCat: let's make HDT scale}  
%
\author{Dennis Diefenbach\inst{1}, Jos\'e M. Gim\'enez-Garc\'ia \inst{1}}
%



\authorrunning{Diefenbach et al.} 
\institute{Univ Lyon, UJM-Saint-\'Etienne, CNRS, Laboratoire Hubert Curien\\
UMR 5516, F-42023 Saint-\'Etienne, France\\
\email{{dennis.diefenbach,jose.gimenez.garcia}@univ-st-etienne.fr}
}

\maketitle              

\begin{abstract}
HDT (Header, Dictionary, Triples) is a serialization for RDF. HDT has become very popular in the last years because it allows to store RDF data with a small disk footprint, while remaining at the same time queriable. For this reason HDT is often used when scalability becomes an issue.

Once RDF data is serialized into HDT, the disk footprint to store it and the memory footprint to query it are very low. However, generating HDT files from raw text RDF serializations (like N-Triples) is a time-consuming and (especially) memory-consuming task. In this publication we present HDTCat, an algorithm and command line tool to join two HDT files with low memory footprint. HDTCat can be used in a divide-and-conquer strategy to generate HDT files from huge datasets using a low-memory footprint.



\end{abstract}

\section{Introduction} \label{sec:introduction}
HDT~\cite{fernandez_binary_2013} is a serialization of RDF like N-triples, Turtle or RDF/JSON to store and exchange RDF. While HDT is not a standard, there exists a corresponding W3C member submission for it\footnote{https://www.w3.org/Submission/HDT/}.\\
Despite other existing RDF formats HDT is characterized by being highly efficient in terms of disk space while being at the same time queriable. It is realistic to compress an RDF file from N-triples to HDT and gain a factor 10 in space. This makes HDT well-suited to exchange large RDF files~\cite{fernandez_binary_2013}. Moreover it is queriable in the sense that by construction one can search for triple patterns. The speed in this case, is comparable to traditional triple-stores~\cite{fernandez_binary_2013}. These two features have made HDT a widely-used technology in the Semantic Web.\\

\subsection{Problem statement}
In the following, we want to briefly describe the problem that is tackled by HDTCat and point out it's importance for HDT in general. Imagine one has two files, file1 and file2 in N-Triples and one wants to merge both files in one file to have all triples together. In a bash script one would do:
\begin{center}
    cat file1 file2 $>$ file1+2
\end{center}
Imagine now that one has two HDT files hdt1 and hdt2 and one wants to merge them, i.e. one wants to create a new HDT file that contains all triples that are contained in hdt1 and hdt2. Merging the two files using "cat" will not create a new HDT file because the structure will not be the right one. So just using "cat" is not a solution.\\
Currently the only solution to merge two HDT files is to convert each HDT file into N-Triples, cat the N-Triples, and generate a new HDT file from it. The problem is that a lot of resources in terms of time and memory are needed.\\
HDT is highly efficient in many aspects, but it's generation is not. For example for compressing 1 Billion N-Triples into HDT 120Gb of RAM are necessary~\cite{gimenez-garcia_hdt-mr:_2015}. So while it is possible to store and query an HDT file containing 1 Billion triples on a modern Laptop, it is not possible to generate the HDT file from another RDF serialization. While time is generally not the bounding factor, memory definitively is. This problem limits the scalability of HDT.\\
This issue is well-known. To address it \cite{gimenez-garcia_hdt-mr:_2015} adapted the HDT generation algorithm such that it can be run using Map-Reduce. This allows to generate HDT files on clusters. While this is a nice solution, it is definitively not handy, since some specific infrastructure is needed.\\
Let's return to the example of merging two HDT files. We observed that the only available solution to merge two HDT files is to decompress them, merge the corresponding N-Triples and generate a new HDT file. The important observation here is that the data-structure of the original HDT files is completely ignored. In this publication we are presenting HDTCat, an algorithm to merge two HDT files in a time and memory efficient way by exploiting the data-structure of HDT. This will not only enable users to easily merge two HDT files, but it represents a significant contribution in the generation of HDT files with low resources, in particular in terms of memory.\\

\subsection{Use of HDT in the Semantic Web Community}
Before we introduce HDTCat we want to point out some works that use HDT.\\
HDT is one of the technologies behind the LOD laundromat~\cite{beek_lod_2014}\footnote{http://lodlaundromat.org/}\footnote{http://lodlaundromat.org/lodlab/}. The LOD laundromat provides and infrastructure that cleans and provides all datasets in the LOD cloud. The infrastructure behind the LOD laundromat takes advantage of HDT thanks to it's small memory footprint.\\
HDT is used by the question answering systems WDAqua-core1\cite{diefenbach_wdaqua-core0:_2017,diefenbach_wdaqua-core1:_2018}. WDAqua-core1 is able to query multiple datasets in the LOD cloud like: Wikidata, DBpedia, MusicBrainz and Dblp. HDT is used as an index for the generation of SPARQL queries and for providing SPARQL endpoints with low space and memory footprint.\\
HDT is also used in the back-end of Linked Data Fragments\footnote{\url{http://linkeddatafragments.org}}, which allows low-cost publication of queryable RDF data by moving intelligence from the server to the client
\cite{verborgh_querying_2014}. HDT is used as the data-structure to resolve triple pattern fragments on the server side. HDT is essential to fullfill the aim of low-cost publication since it allows to resolve triple patterns in a rapid way with small resource consumption.\\
Finally HDT is used by a command line tool called PageRankRDF \footnote{\url{https://github.com/WDAqua/PageRankRDF}} to compute PageRank scores over RDF datasets~\cite{diefenbach_pagerank_2018}. PageRank computations are characterized by high memory footprint. The compression of HDT allows to reduce this footprint significantly.\\

To sum up, HDT represents one of the key technologies in the Semantic Web and is often used when scalability becomes a problem. At the same time generating HDT files from existing RDF serializations is a memory intense task and scalability is a big issue. We want to address this issue in the following by introducing HDTCat.

\section{Background} \label{sec:background}

In this section we provide basic background knowledge about RDF and how it is serialized into HDT. This is necessary to understand the approach to merge two HDT files.

\subsection{RDF} \label{subsec:background.rdf}

RDF is the data model used in the Semantic Web. The data is organized in \emph{triples} in the form \triple, where \s (the subject) is the resource being described, \p (the predicate) is the property that describes it, and \o (the object) is the actual value of the property. An object can be either a resource or a literal value. In a set of triples, resources can appear as subject or object in different triples, forming a directed labeled graph, which is known as \emph{RDF graph}. Formal definitions for RDF triple and RDF graph (adapted from \cite{gutierrez_foundations_2011}) can be seen in Definition \ref{def:RDFTriple} and \ref{def:RDFGraph}, respectively. 

\begin{definition}[RDF triple]\label{def:RDFTriple}
Assume an infinite set of terms $\terms = \I \cup \B \cup \L$, where \I, $\B$, and \L are mutually disjoint, and \I are IRI references, \B are Blank Nodes, and \L are Literals. An RDF triple is a tuple $\triple \in (\I \cup \B) \times \I \times (\I \cup \B \cup \L)$, where ``\s'' is the subject, ``\p'' is the predicate and ``\o'' is the object. 
\end {definition}

\begin{definition}[RDF graph\label{def:RDFGraph}]
An RDF graph \G is a set of RDF triples of the form $\triple$. It can be represented as a directed labeled graph whose edges are $\s\xrightarrow{\p} \o$. 
\end{definition}

\begin{example}\label{rdf}
The following snippet show an RDF file, that we call $RDF_{1}$, in N-Triples format:
\begin{lstlisting}[language=N3]
<so1> <p1> <o1> .
<so1> <p1> <o2> .
<s1>  <p2> <so1> .
\end{lstlisting}
The following RDF file in N-Triples, we denote as $RDF_{2}$:
\begin{lstlisting}[language=N3]
<so1> <p3> <o2> .
<o2> <p1> <s1> .
\end{lstlisting}
We will use these two files as running examples and show how they can be compressed and merged using HDTCat.
\end{example}

\subsection{HDT} \label{subsec:background.hdt}

HDT~\cite{fernandez_binary_2013} is a binary serialization format for RDF based in compact data structures. Compact data structures are data structures that compress the data as close as possible to its theoretic lower bound, but allow for efficient query operations. HDT encodes an RDF graph as a set of three components: (1) Header, (2) Dictionary, and (3) Triples.

The \emph{Header} component stores metadata information about the RDF dataset and the HDT serialization itself. This data can be necessary to read the other sections of an HDT file. The \emph{Dictionary} component stores the different IRIs, blank nodes, and literals, and assigns to each one an unambiguous integer ID. The \emph{Triples} component stores the RDF graph, where all the terms are replaced by the ID assigned in the Dictionary component. From now on to represent an HDT file, we write $\HDT=(\H,\D,\T)$, where \H is the header component, \D is the dictionary component, and \T is the triples component. In theory, each component allows different encodings. In practice, however, current compression formats are based in sorting lexicographically their elements. We describe thereafter characteristics of current HDT encodings.

The header is stored in raw text, allowing consumers to read it without any additional knowledge or tools. The data is not compressed.
In the dictionary an integer ID is assigned to each term (IRI, Blank Node and Literal). The set of terms is divided in four sections: (1) the \emph{Shared} section, that stores the terms that appear at the same time as subjects and objects of triples; the \emph{Subjects} section, which stores the terms that appear exclusively as subjects of triples; the \emph{Objects} section, which contains the terms that appear only as object of triples; and finally the \emph{Predicates} section, storing the terms that appear as predicates of the triples. From now on, we write the Dictionary as a tuple $\D=(\SO,\S,\O,\P)$, where \SO is the shared section, \S is the subjects section, \O is the objects section, and \P is the predicates section.
In each section the terms are sorted lexicographically and compressed (\emph{e.g.}, using Plain~\cite{brisaboa_compressed_2011} or Hu-Tucker FrontCoding~\cite{hu_optimal_1971}). The position of each term is then used as its implicit ID in each section. This way to each term an integer is assigned in a space efficient way. The dictionary needs to provide global IDs for subjects and objects, independently of the section in which they are stored. Terms in \P and \SO do not change, while IDs for \S and \O sections are increased by the size of \SO (\textit{i.e.}, $\ID[\s]=\ID[\S[\s]]+\max(\ID[\SO])$ and $\ID[\o]=\ID[\O[\o]]+\max(\ID[\SO])$).\\

\begin{example}\label{example}
Consider the file $RDF_{1}$ from example \ref{rdf}. We call the corresponding HDT file $\HDT[1]=(\H[1],\D[1],\T[1])$ with $\D[1]=(\SO[1],\S[1],\O[1],\P[1])$. The dictionary sections look as follows (note that the compression is not shown here as it is not important to understand HDTCat):
\begin{center}
\begin{tabular}{ll}
  \multicolumn{2}{c}{\SO[1]} \\
  \hline
  \multicolumn{1}{|c}{IRI} & \multicolumn{1}{|c|}{ID}  \\
  \hline
  \multicolumn{1}{|c}{\textless so1\textgreater}& \multicolumn{1}{|c|}{1}\\
  \hline
  &  \\
\end{tabular}
\quad
\begin{tabular}{ll}
  \multicolumn{2}{c}{\S[1]} \\
  \hline
  \multicolumn{1}{|c}{IRI} & \multicolumn{1}{|c|}{ID}  \\
  \hline
  \multicolumn{1}{|c}{\textless s1\textgreater}& \multicolumn{1}{|c|}{2}\\
  \hline
  &  \\
\end{tabular}
\quad
\begin{tabular}{ll}
  \multicolumn{2}{c}{\O[1]} \\
  \hline
  \multicolumn{1}{|c}{IRI} & \multicolumn{1}{|c|}{ID}  \\
  \hline
  \multicolumn{1}{|c}{\textless o1\textgreater}& \multicolumn{1}{|c|}{2}\\
  \hline
  \multicolumn{1}{|c}{\textless o2\textgreater}& \multicolumn{1}{|c|}{3}\\
  \hline
\end{tabular}
\quad
\begin{tabular}{|l|l|}
  \multicolumn{2}{c}{\P[1]} \\
  \hline
  \multicolumn{1}{|c}{IRI} & \multicolumn{1}{|c|}{ID}  \\
  \hline
  \textless p1\textgreater & 1\\
  \textless p2\textgreater & 2\\
  \hline
\end{tabular}
\end{center}
Note that the ids in the \S[1] and \O[2] section start by $2$ since there is one entry in the common section \SO[1]. Similarly for $RDF_{2}$ we get $HDT_{2}$ with:
\begin{center}
\begin{tabular}{ll}
  \multicolumn{2}{c}{SO\textsubscript{2}} \\
  \hline
  \multicolumn{1}{|c}{IRI} & \multicolumn{1}{|c|}{ID}  \\
  \hline
  \multicolumn{1}{|c}{\textless o2\textgreater}& \multicolumn{1}{|c|}{1}\\
  \hline
  \multicolumn{1}{|c}{\textless so1\textgreater}& \multicolumn{1}{|c|}{2}\\
  \hline
\end{tabular}
\quad
\begin{tabular}{ll}
  \multicolumn{2}{c}{\S[2]} \\
  \hline
  \multicolumn{1}{|c}{IRI} & \multicolumn{1}{|c|}{ID}  \\
  \hline
  &  \\
  &  \\
\end{tabular}
\quad
\begin{tabular}{ll}
  \multicolumn{2}{c}{\O[2]} \\
  \hline
  \multicolumn{1}{|c}{IRI} & \multicolumn{1}{|c|}{ID}  \\
  \hline
  \multicolumn{1}{|c}{\textless s1\textgreater}& \multicolumn{1}{|c|}{3}\\
  \hline
  &  \\
\end{tabular}
\quad
\begin{tabular}{|l|l|}
  \multicolumn{2}{c}{\P[2]} \\
  \hline
  \multicolumn{1}{|c}{IRI} & \multicolumn{1}{|c|}{ID}  \\
  \hline
  \textless p1\textgreater & 1\\
  \textless p3\textgreater & 2\\
  \hline
\end{tabular}
\end{center}

\begin{flushright}
$\square$
\end{flushright}
\end{example}

In the triples component, each term  in the triples is replaced by its ID from the dictionary and sorted in what is known as \emph{Plain Triples}. The ordering is defined in the following.
\begin{definition}\label{def:ordering}
If $\T[1]=(\s[1],\p[1],\o[1])$ and $\T[2]=(\s[2],\p[2],\o[2])$ are two triples then $\T[1]\geq \T[2]$ if and only if:
\begin{enumerate}
    \item $\s[1]\geq \s[2]$;
    \item if $\s[1]=\s[2]$ then $\p[1]\geq \p[2]$;
    \item if $\s[1]=\s[2]$ and $\p[1]=\p[2]$ then $\o[1]\geq \o[2]$;
\end{enumerate}
\end{definition}

\begin{example}
The triples from $RDF_{1}$ from Example \ref{example} in \emph{Plain Triples} are:
\begin{lstlisting}[language=N3]
1 1 2 
1 1 3
2 2 1
\end{lstlisting}
Note that they respect the order defined in Definition \ref{def:ordering}. The one from $RDF_{2}$ are:
\begin{lstlisting}[language=N3]
1 1 3
2 2 1
\end{lstlisting}
Note that the triples were reordered.
\begin{flushright}
$\square$
\end{flushright}
\end{example}
The triples can be compressed in \emph{Compact Triples}, which uses two coordinated sequences of IDs, \Seq[\P] and \Seq[\O], to store the IDs of predicates and objects respectively, in the order they appear in the sorted triples. The first ID in \Seq[\P] is assumed to have the subject with $\ID[\s]=1$. Each following ID is assumed to have the same ID as its predecessor. If the ID $0$ appears in the sequence, it means a change to the following ID (\textit{i.e.}, the ID is incremented by one). Respectively, the first ID in \Seq[\O] is matched with the property in the first position of \Seq[\P]. Each following ID is assumed to have the property as its predecessor, and if the ID $0$ appears in the sequence, it means a change to the following ID (that is, the next ID in \Seq[P]). This can be further compressed in \emph{BitMap Triples} by removing the $0$ from the ID sequences and adding two bit sequences, \Bit[\P] and \Bit[\O], that mark the position where the change of subject (for \Seq[\P]) or predicate (for \Seq[\O]) happen.\\
Note that the data-structures described above allow fast retrieval of all triple patterns with fixed subject. Some more indexes are added to resolve fast triple patterns with fixed predicate or object. Moreover note that due to the global ordering updates are not supported.

\section{HDTCat} \label{sec:hdtcat}

In this section we are going to describe the algorithm behind HDTCat. Our goal is, given two HDT files \HDT[1] and \HDT[2], to generate a new HDT file called \HDT[cat] that contains the union of the triples in \HDT[1] and \HDT[2].\\
The goal is to achieve this in a scalable way in particular in terms of memory footprint since this is generally the limited resource on current hardware.\\
The algorithm can be decomposed into three phases:
\begin{itemize}
    \item phase 1: joining the dictionaries,
    \item phase 2: joining the triples,
    \item phase 3: generate the header.
\end{itemize}

\begin{algorithm}[t]
 \KwData{Two sorted lists a and b}
 \KwResult{A sorted list c containing all entities in a and b }
 n= length of a; m = length of b \\
 allocate c with length n+m \\
 i = 1; j = 1 \\
 \While{$i < n$ $||$ $j < m$}{
    \If{$i = n$}{
        copy rest of b into c\\
        break\\
    }
    \If{$j = m$}{
        copy rest of a into c\\
        break\\
    }
    \If{$a[i] < b[j]$}{
        copy a[i] into c\\
        i=i+1\\
    }
    \If{$b[j] < a[i]$}{
        copy b[j] into c\\
        j=j+1\\
    }
    \If{$a[i] = b[j]$}{
    copy a[i] into c\\
    i=i+1\\
    j=j+1\\
    }
 }
\caption{Algorithm to merge two sorted lists. Note that the algorithm has a time complexity of $O((n+m))$. All computation do not need to be done on RAM but can be performed on disk.}
\label{fig:algo1}
\end{algorithm}

Let's assume two HDT files, $\HDT[1]=(\H[1],\D[1],\T[1])$ and $\HDT[2]=(\H[1],\D[1],\T[1])$ are given. The problem is how to merge the dictionaries \D[1], \D[2] and the triples \T[1], \T[2] so that the resulting HDT file contains the union of the RDF triples.\\
The main idea is the following. Both the dictionary and the triples are basically sorted lists.\\
Remember, the current solution to merge two HDT files does the following: first the HDT files are decompressed, the resulting triples are cat to one file and compresses again. Basically two ordered lists are put one after the other and ordered again without exploiting the initial order of the lists.\\
The algorithm behind HDTCat is taking into consideration the ordering of the lists. The lists are merged by the algorithm described in Figure \ref{fig:algo1}. It does the following. There are two iterators over the two lists. Recursively the current entries of the two iterators are compared and the lowest entry is added to the final list.\\
There are two important consequences. Imagine the two lists have $n$ respectively $m$ entries. The first consequence is that the time complexity is reduced. When HDT is compressed, the current implementation uses merge sort, which has a time complexity of $O((n+m)\cdot log(n+m))$. When sorting two already sorted lists, using the algorithm above, the time complexity is $O(n+m)$.\\
The second, and in our eyes the more important, is the memory consumption. The existing approach stores every uncompressed triple in memory so that the memory complexity is in the order of O(n+m). Iterating over the sorted lists by letting them compressed and decompressing only the current entry reduced dramatically the memory footprint.\\
This explains the main idea behind HDTCat. We are now going to explain more in detail the merging strategy and the data-structures needed.

\begin{figure}[t]
\begin{center}
\includegraphics[width=0.5\textwidth]{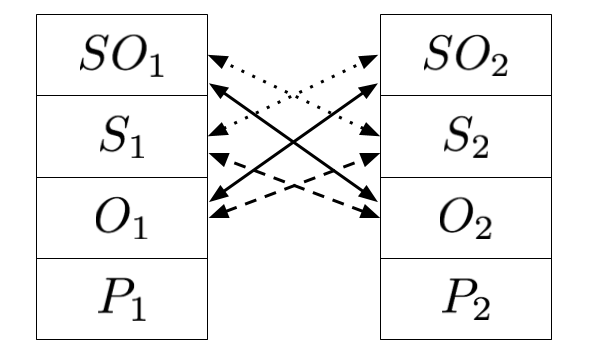}
\end{center}
\caption{This figure shows the non-trivial sections that can share an entry. Clearly \SO[1] and \SO[2], \S[1] and \S[2], \O[1] and \O[2], \P[1] and \P[2] can contain common entries. The other sections that can contain common entries are connected by a double arrow. It is important to take care of these common entries when merging the dictionaries.}
\label{fig:move1}
\vspace{-5mm}
\end{figure}

\subsection{Joining the dictionary}
Assume two HDT dictionaries $\D[1]=(\SO[1],\S[1],\O[1],\P[1])$ and $\D[2]=(\SO[2],\S[2],\O[2],\P[2])$ are given. We want to create the new HDT dictionary $\D[\cat]=(\SO[\cat],\S[\cat],\O[\cat],\P[\cat])$.\\
Merging the section \P[1] and \P[2] is easy. \P[1] and \P[2] are two arrays of ordered compressed strings. Algorithm \ref{fig:algo1} assumes that there are two iterators over the two lists. Recursively the current entries of the two iterators are compared and the lowest entry is added to the final list. To compare the entries these are decompressed and the new entry is compressed directly and added to \P[\cat].
Note that since the strings are uncompressed and compressed directly, the memory footprint remains low.
\begin{example}
The predicate section of \HDT[\cat] is:\\
\begin{tabular}{|l|l|}
  \multicolumn{2}{c}{\P[\cat]} \\
  \hline
  IRI & ID  \\
  \hline
  \textless p1\textgreater & 1 \\
  \textless p2\textgreater  & 2 \\
  \textless p3\textgreater  & 3 \\
  \hline
\end{tabular}
\begin{flushright}
$\square$
\end{flushright}
\end{example}
Merging the other sections is as easy, with the exception that ids can move from one section to another. For example if \S[1] contains a IRI that appears also in \O[2] then the corresponding entry must be moved to the section \SO[\cat] (since the IRI will appear both in the subject and the object of some triples). Figure \ref{fig:move1} shows the sections that can contain common elements (excluding the non-trivial cases). One must take care of the following cases:
\begin{itemize}
    \item If \SO[1] and \S[2], or \S[1] and \SO[2] contain common entries, then they must be skipped when joining the $S$ sections.
    \item If \SO[1] and \O[2], or \O[1] and \SO[2] contain common entries, then they must be skipped when joining the $O$ sections.
    \item If \S[1] and \O[2], or \O[1] and \S[2] contain common entries then they must be skipped when joining the \S and \O sections, and additionally they must be added to the \SO[\cat] section.
\end{itemize}

\begin{figure}[t]
\begin{center}
\includegraphics[width=0.5\textwidth]{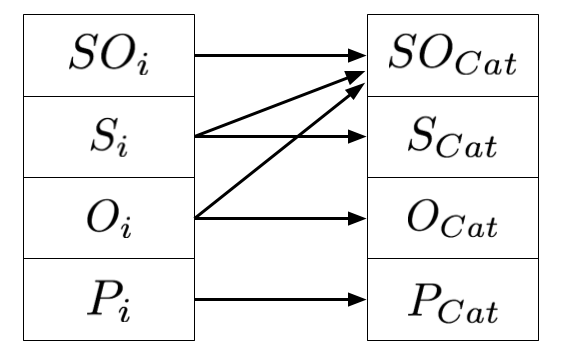}
\end{center}
\caption{This figure shows to which sections of the \HDT[\cat] dictionary, the entries from the dictionary section of either \HDT[1] or \HDT[2] can move. The \SO section and the \P section ids are going to the \SO[\cat] and \P[\cat] section respectively. If there is an entry that appears both in the \S section and the \O section, then the corresponding entry will go to the \SO[\cat] section. Otherwise the entry goes to the \S or \O section.}
\label{fig:move2}
\vspace{-5mm}
\end{figure}

In particular entries can move from one section to another. Figure \ref{fig:move2} shows to which sections the entries can move. This describes how the sections of the dictionary are merged. We modified the existing data-structures to store the dictionary sections of \HDT[\cat] directly on disk.

\begin{example}
The sections of \HDT[\cat] different from \P[\cat] look like this:
\begin{center}
\begin{tabular}{|l|l|}
  \multicolumn{2}{c}{\SO[\cat]} \\
  \hline
  IRI & ID  \\
  \hline
  \textless o2\textgreater & 1 \\
  \textless so1\textgreater & 2 \\
  \textless s1\textgreater  & 3 \\
  \hline
\end{tabular}
\quad
\begin{tabular}{ll}
  \multicolumn{2}{c}{\S[\cat]} \\
  \hline
  \multicolumn{1}{|c|}{IRI} & \multicolumn{1}{|c|}{ID}  \\
  \hline
  & \\
  &  \\
  &  \\
  
\end{tabular}
\quad
\begin{tabular}{ll}
  \multicolumn{2}{c}{\O[\cat]} \\
  \hline
  \multicolumn{1}{|c|}{IRI} & \multicolumn{1}{|c|}{ID}  \\
  \hline
  \multicolumn{1}{|c}{\textless o1\textgreater} & \multicolumn{1}{|c|}{4} \\
  \hline
  &  \\
  &  \\
\end{tabular}
\end{center}
Note that the IRI \textless s1\textgreater moved from section \S[1] to section \SO[\cat]. 
\begin{flushright}
$\square$
\end{flushright}
\end{example}

When joining the triples we need to know what is the correspondence between the ids in \D[1] and \D[2] and the ones in \D[\cat]. To keep track where the ids are mapped, we introduce data structures which assign for each ID in the sections $\Sec\in\{\SO[1],\S[1],\O[1],\P[1],\SO[2],\S[2],\O[2],\P[2]\}$ the new ID in the sections $\Sec[\cat]\in\{\SO[\cat],\S[\cat],\O[\cat],\P[\cat]\}$. For one section \Sec the data structure contains two arrays:
\begin{itemize}
    \item an array indicating for each ID of \Sec which is the corresponding section in \Sec[\cat].   
    \item an array mapping the ids of \Sec to the corresponding ID in the section \Sec[\cat]. 
\end{itemize}
We indicate every such mapping as \mapping{\Sec}. Moreover we construct also the mappings form  
\SO[\cat],\S[\cat] (note: the IDs of these two sections are consecutive) to \SO[1],\S[1] and \SO[2],\S[2] respectively. This consists of two arrays:
\begin{itemize}
    \item an array indicating for each ID of \SO[\cat],\S[\cat] the corresponding ID in \SO[1],\S[1] (if it exists).   
    \item an array indicating for each ID of \SO[\cat],\S[\cat] the corresponding ID in \SO[2],\S[2] (if it exists).
\end{itemize}
The arrays are directly written to disk. We indicate the two mapping as \mapping[1]{\cat} and \mapping[2]{\cat}. 
\begin{example}\label{mapping}
The mappings for HDTCat are as follows:
\begin{center}
\begin{tabular}{lll}
  \multicolumn{3}{c}{\mapping{\SO[1]}} \\
  \hline
  \multicolumn{1}{|c}{ID} & \multicolumn{1}{|c}{Sec\textsubscript{\cat}} & \multicolumn{1}{|c|}{ID\textsubscript{\cat}}   \\
  \hline
  \multicolumn{1}{|c}{1}& \multicolumn{1}{|c}{\SO[\cat]} & \multicolumn{1}{|c|}{2}\\
  \hline
  &  \\
  &  \\
\end{tabular}
\quad
\begin{tabular}{lll}
  \multicolumn{3}{c}{\mapping{\S[1]}} \\
  \hline
  \multicolumn{1}{|c}{ID} & \multicolumn{1}{|c}{Sec\textsubscript{\cat}} & \multicolumn{1}{|c|}{ID\textsubscript{\cat}}   \\
  \hline
  \multicolumn{1}{|c}{2}& \multicolumn{1}{|c}{\SO[\cat]} & \multicolumn{1}{|c|}{3}\\
  \hline
  &  \\
  &  \\
\end{tabular}
\quad
\begin{tabular}{lll}
  \multicolumn{3}{c}{\mapping{\O[1]}} \\
  \hline
  \multicolumn{1}{|c}{ID} & \multicolumn{1}{|c}{Sec\textsubscript{\cat}} & \multicolumn{1}{|c|}{ID\textsubscript{\cat}}   \\
  \hline
  \multicolumn{1}{|c}{2}& \multicolumn{1}{|c}{\O[\cat]} & \multicolumn{1}{|c|}{4}\\
  \hline
  \multicolumn{1}{|c}{3}& \multicolumn{1}{|c}{\SO[\cat]} & \multicolumn{1}{|c|}{1}\\
  \hline
  &  \\
\end{tabular}
\quad
\begin{tabular}{lll}
  \multicolumn{3}{c}{\mapping{\P[1]}} \\
  \hline
  \multicolumn{1}{|c}{ID} & \multicolumn{1}{|c}{Sec\textsubscript{\cat}} & \multicolumn{1}{|c|}{ID\textsubscript{\cat}}   \\
  \hline
  \multicolumn{1}{|c}{1}& \multicolumn{1}{|c}{\P[\cat]} & \multicolumn{1}{|c|}{1}\\
  \hline
  \multicolumn{1}{|c}{2}& \multicolumn{1}{|c}{\P[\cat]} & \multicolumn{1}{|c|}{2}\\
  \hline
  &  \\
\end{tabular}
\begin{tabular}{lll}
  \multicolumn{3}{c}{\mapping{\SO[2]}} \\
  \hline
 \multicolumn{1}{|c}{ID} & \multicolumn{1}{|c}{Sec\textsubscript{\cat}} & \multicolumn{1}{|c|}{ID\textsubscript{\cat}}   \\
  \hline
  \multicolumn{1}{|c}{1}& \multicolumn{1}{|c}{\SO[\cat]} & \multicolumn{1}{|c|}{1}\\
  \hline
  \multicolumn{1}{|c}{2}& \multicolumn{1}{|c}{\SO[\cat]} & \multicolumn{1}{|c|}{2}\\
  \hline
  &  \\
\end{tabular}
\quad
\begin{tabular}{lll}
  \multicolumn{3}{c}{\mapping{\S[2]}} \\
  \hline
  \multicolumn{1}{|c}{ID} & \multicolumn{1}{|c}{Sec\textsubscript{\cat}} & \multicolumn{1}{|c|}{ID\textsubscript{\cat}}   \\
  \hline
  &  \\
  &  \\
  &  \\
\end{tabular}
\quad
\begin{tabular}{lll}
  \multicolumn{3}{c}{\mapping{\O[2]}} \\
  \hline
  \multicolumn{1}{|c}{ID} & \multicolumn{1}{|c}{Sec\textsubscript{\cat}} & \multicolumn{1}{|c|}{ID\textsubscript{\cat}}   \\
  \hline
  \multicolumn{1}{|c}{3}& \multicolumn{1}{|c}{\SO[\cat]} & \multicolumn{1}{|c|}{1}\\
  \hline
  &  \\
  &  \\
\end{tabular}
\quad
\begin{tabular}{lll}
  \multicolumn{3}{c}{\mapping{\P[2]}} \\
  \hline
  \multicolumn{1}{|c}{ID} & \multicolumn{1}{|c}{Sec\textsubscript{\cat}} & \multicolumn{1}{|c|}{ID\textsubscript{\cat}}   \\
  \hline
  \multicolumn{1}{|c}{1}& \multicolumn{1}{|c}{\P[\cat]} & \multicolumn{1}{|c|}{1}\\
  \hline
  \multicolumn{1}{|c}{2}& \multicolumn{1}{|c}{\P[\cat]} & \multicolumn{1}{|c|}{3}\\
  \hline
  &  \\
\end{tabular}

\begin{tabular}{lll}
  \multicolumn{2}{c}{\mapping[1]{\cat}} \\
  \hline
  \multicolumn{1}{|c}{ID\textsubscript{\cat}} & \multicolumn{1}{|c|}{ID\textsubscript{old}}   \\
  \hline
  \multicolumn{1}{|c}{1} & \multicolumn{1}{|c|}{-}\\
  \hline
  \multicolumn{1}{|c}{2} & \multicolumn{1}{|c|}{1}\\
  \hline
  \multicolumn{1}{|c}{3} & \multicolumn{1}{|c|}{2}\\
  \hline
\end{tabular}
\quad
\begin{tabular}{lll}
  \multicolumn{2}{c}{\mapping[2]{\cat}} \\
  \hline
  \multicolumn{1}{|c}{ID\textsubscript{cat}} & \multicolumn{1}{|c|}{ID\textsubscript{old}}   \\
  \hline
  \multicolumn{1}{|c}{1} & \multicolumn{1}{|c|}{1}\\
  \hline
  \multicolumn{1}{|c}{2} & \multicolumn{1}{|c|}{2} \\
  \hline
  \multicolumn{1}{|c}{3} & \multicolumn{1}{|c|}{-}\\
  \hline
\end{tabular}

\end{center}
\begin{flushright}
$\square$
\end{flushright}
\end{example}
This describes how the dictionaries are merged by HDTCat.

\subsection{Joining the triples}
We now want to describe how to join the triples \T[1], \T[2]. To join the triples we are only indirectly exploiting the fact that the triples are ordered, namely by the fact that the HDT files are queriable.\\
Remember that by Definition \ref{def:ordering} we have to order the triples first by subjects, then by predicates and finally by objects. So we have first to order the ids in \S[\cat]. This is trivial, we already know, when constructing the dictionary, how many subjects exist and since they are integers they are sorted. We now need, for every ID in \S[\cat], to find and order all the associated triples. So assume we want to find and order all triples with the ID \ID[\cat]. We use the mappings $M(S[\cat],S[1])$ and $M(S[\cat],S[2])$, constructed when joining the dictionary sections, to find the ids that mapped to \ID[\cat]. Assume that \ID[1] and \ID[2] are these ids. Since both \HDT[1] and \HDT[2] are queriable, we can retrieve all triples with subjects \ID[1] and \ID[2] respectively. By using again the mappings constructed when joining the dictionaries, we can now translate the ids of these triples used in \HDT[1] and \HDT[2] to the corresponding IDs in \HDT[\cat]. It is very important to note that the triples associated to \ID[1] and \ID[2] in \HDT[1] and \HDT[2]  are ordered, but this ordering is lost after mapping them because some of the ids move from one section to another! It is therefore not possible to use Algorithm \ref{fig:algo1} for sorting, but some standard sorting algorithm, like merge sort, is needed. Finally note that this is still a big computational advantage because instead of ordering all triples, we order many sub-lists, which is computationally better in terms of time and memory. We generate the triples by iterating over the subjects and by writing the triples directly to disk.\\
\begin{example}
Let's first join the triples with $\ID[\cat]=1$. According to $M_{Cat,2}$ there are only triples in \HDT[2] mapping to it. In fact there is only the triple:\\
1  2  3\\
By using the mappings of Example \ref{mapping} this will become:\\
1  3  1\\
For $\ID[\cat]=2$ we search all triples associated to $\ID[\cat]=2$. These triples are:\\
1   1   2\\
1   1   3\\
in \HDT[1] and:\\
2   2   1\\
in \HDT[2]. By using the mappings of Example \ref{mapping} these correspond to the new IDs:\\
2   1   3\\
2   1   1\\
and:\\
2   3   1\\
Note that the triples of \HDT[1] where initially ordered, while the mapped triples are not ((2,1,3)$>$(2,1,1)). The merged triples for $\ID[\cat]=2$ are then:\\
2   1   1\\
2   1   3\\
2   3   1
\end{example}
\begin{flushright}
$\square$
\end{flushright}
This explains how the triple components are merged by HDTCat.

\subsection{Creating the Header}
While the dictionary and the triples must be merged from the corresponding sections of the two HDT files, the header just contain some statistical information like the number of triples and the number of distinct subjects. This means that there is nothing to do here except writing the statistics corresponding to \D[\cat] and \T[\cat] that were generated.

\section{Experiment}
In this section we analyze the performance of HDTCat. In particular we analyze how HDTCat solves the memory scalability problem when generating HDT files starting from other RDF serializations like ntriples.\\ 
There are 3 methods to compute HDT files from other RDF serializations namely: the command line tool \tttt{rdf2hdt} that is part of the HDT repository\footnote{\url{https://github.com/rdfhdt/hdt-java}}, HDT-MR\cite{gimenez-garcia_hdt-mr:_2015} and HDTCat.\\

\textbf{Experiment 1} We compare the 3 methods to generate HDT from other RDF sterilizations by using the syntetic data provided by LUBM \cite{guo_lubm:_2005}. LUBM is a benchmark to test the performance of SPARQL queries and contains both a tool to generate syntetic RDF data and a set of SPARQL query. The generated RDF contains information about universities like departments, students, professors and so on. We generated the following LUBM datasets: (1) from 1000 to 8000 universities in steps of 1000, and (2) from 8000 to 40000 universities in steps of 4000. We used the 3 methods in the following way:

\begin{table}
\centering
\caption{Comparison between methods to serialize RDF into HDT.}
\label{table-result}
\scalebox{0.9}{
\begin{tabular}{|l|l|l|l|l|l|l|l|l|l|}
\hline
\multicolumn{9}{|c|}{\textbf{Configuration 1: 128 Gb RAM}}\\
\hline
\multirow{2}{*}{LUBM} & \multirow{2}{*}{Triples} & \multicolumn{2}{c}{hdt2rdf} & \multicolumn{1}{|c}{HDT-MR} & \multicolumn{4}{|c|}{HDTCat}   \\
\cline{3-9}
     & & T (s) & M (Gb) & T (s) & T\_com (s) & T\_cat (s) & T (s) & M\_cat (Gb) \\
\hline
1000  & 0.13BN & 1856  & 53.4  &  936   & \multirow{8}{*}{$970^{\ast}$}  & - & - & - \\
2000  & 0.27BN & 4156  & 70.1  &  1706  &  & 317  & 2257 & 26.9 \\
3000  & 0.40BN & 6343  & 89.3  &  2498  &  & 468  & 3695 & 35.4 \\
4000  & 0.53BN & 8652  & 105.7  &  3113  &  & 620  & 5285 & 33.8 \\
5000  & 0.67BN & 11279  & 118.9  &  4065  &  & 803  & 7058 & 41.7 \\
6000  & 0.80BN & 23595  & 122.7  &  4656  &  & 932  & 8960 & 47.5 \\
7000  & 0.93BN & 78768  & 123.6  &  5338  &  & 1088 & 11018 & 52.9 \\
8000  & 1.07BN & $\star$  & $\star$ &  6020  &  & 1320 & 13308 & 58.7 \\
\cline{6-6}
12000 & 1.60BN & - & - &  9499  & \multirow{8}{*}{$4710^{\ast}$} & 1759 & 19777 & 54.7 \\
16000 & 2.14BN & - & - &  13229 &  & 2338 & 26825 & 73.4  \\
20000 & 2.67BN & - & - &  15720 &  & 2951 & 34486 & 90.5 \\
24000 & 3.20BN & - & - &  26492 &  & 3593 & 42789 & 90.6 \\
28000 & 3.74BN & - & - &  36818 &      & 4308 & 51807 & 84.9 \\
32000 & 4.27BN & - & - &  40633 &      & 4849 & 61366 & 111.1 \\
36000 & 4.81BN & - & - &  48322 &      & 6085 & 72161 & 109.4 \\
40000 & 5.32BN & - & - &  55471 &      & 7762 & 84633 & 100.1\\
\hline


\multicolumn{9}{|c|}{\textbf{Configuration 2: 32 Gb RAM}}\\
\hline
\multirow{2}{*}{LUBM} & \multirow{2}{*}{Triples} & \multicolumn{2}{c|}{HDT} & - & \multicolumn{4}{c|}{HDTCat}   \\
\cline{3-9}
     & & T (s) & M (Gb) & - & T\_com (s) & T\_cat (s) & T (s) & M\_cat (Gb) \\
\hline
1000  & 0.13BN & 1670 & 28.3  & - & \multirow{8}{*}{$1681^{\ast}$} & -  & - & -\\
2000  & 0.27BN & $\star$  & $\star$  & - &  & 454  & 3816 & 17.3\\
3000  & 0.40BN & - & - & - &  & 660  & 6366 & 20.1\\
4000  & 0.53BN & - & - & - &  & 869 & 8916 & 25.5\\
5000  & 0.67BN & -  & - & - &  & 1097 & 11694 &29.3\\
6000  & 0.80BN & -  & - & - &  & 1345 & 14720 &28.5\\
7000  & 0.93BN & -  & - & - &  & 1584 & 17985 &30.6\\
8000  & 1.07BN & -  & - & - &  & 1830 & 21496 &30.4\\
\cline{6-6}
12000 & 1.60BN & - & - & - & $\star$  & 2748 & - & 31.0\\
\cline{6-6}
16000 & 2.14BN & - & - & - & - & 3736   & -  & 31.1\\
20000 & 2.67BN & - & - & - & - & 5007 & - & 30.5\\
24000 & 3.20BN & - & - & - & - & 5514 & - & 30.8\\
28000 & 3.74BN & - & - & - & - & 6568 & - & 30.8\\
32000 & 4.27BN & - & - & - & - & 7358 & - & 30.8\\
36000 & 4.81BN & - & - & - & - & 9126 & - & 30.6\\
40000 & 5.32BN & - & - & - & - & 9711 & - & 30.8\\
\hline

\multicolumn{9}{|c|}{\textbf{Configuration 3: 16 Gb RAM}}\\
\hline
\multirow{2}{*}{LUBM} & \multirow{2}{*}{Triples} & \multicolumn{2}{c}{HDT} & - & \multicolumn{4}{|c|}{HDTCat}   \\
\cline{3-9}
     & & T (s) & M (Gb) & - & T\_com (s) & T\_cat (s) & T (s) & M\_cat (Gb) \\
\hline
1000  & 0.13BN & 2206  &  14.5 & - & \multirow{8}{*}{2239*} & - & - & -\\
2000  & 0.27BN & $\star$ & $\star$ & - &  &  517 & 4995 & 10.7\\
3000  & 0.40BN & - & - & - &  &  848 & 8082 &  11.8\\
4000  & 0.53BN & - & - & - &  &  1301 & 11622 & 11.9\\
5000  & 0.67BN & - & - & - &  &  1755 & 15616 & 12.7\\
6000  & 0.80BN & - & - & - &  &  2073 & 19928 & 11.8\\
7000  & 0.93BN & - & - & - &  &  2233 & 24400 & 12.6\\
8000  & 1.07BN & - & - & - &  &  3596 & 30235 & 12.2\\
\cline{6-6}
12000 & 1.60BN & - & - & - & $\star$ & 4736 & -  & 14.3\\
\cline{6-6}
16000 & 2.14BN & - & - & - & - & 6640 & - & 14.3\\
20000 & 2.67BN & - & - & - & - & 9058 & - & 14.4\\
24000 & 3.20BN & - & - & - & - & 10102 & - & 14.3\\
28000 & 3.74BN & - & - & - & - & 13287 & - & 12.8\\
32000 & 4.27BN & - & - & - & - & 14001 & - & 13.9\\
36000 & 4.81BN & - & - & - & - & 17593 & - & 14.0\\
40000 & 5.32BN & - & - & - & - & 19929 & - & 13.9\\
\hline
\end{tabular}
}
\end{table}

\begin{itemize}
    \item \textbf{rdf2hdt} We cat the above datasets to obtain the following datasets lubm.1-1.000.nt, lubm.1-2.000.nt, ..., lubm.1-8.000.nt and lubm.1-12.000.nt, lubm.1-16.000.nt, ...., lubm.1-40.000.nt. We then used rdf2hdt to compute the corresponding HDT files. So we first cat the datasets and then converted them in HDT.
    \item \textbf{HDT-MR} HDT-MR is used in the same way as rdf2hdt. First, the datasets are cat and then converted to HDT. 
    \item \textbf{HDTCat} We first computed the HDT datasets lubm.1-1.000.hdt, lubm.1-2.000.hdt, ..., lubm.1-8.000.hdt and lubm.8.001-12.000.hdt, lubm.12.0001-16.000.hdt, ...., lubm.36.000-40.000.hdt using rdf2hdt. Then we used HDTCat to recursively compute lubm.1-1.000.hdt, lubm.1-2.000.hdt, ..., lubm.1-8.000.hdt and lubm.1-12.000.hdt, lubm.1-16.000.hdt, ...., lubm.1-40.000.hdt. So we first generate the HDT files and we then cat them toghether using HDTCat
\end{itemize}

We executed the experiments for rdf2hdt and HDTCat on different hardware configurations:
\begin{itemize}
    \item \textbf{Configuration 1}: A server with 128 Gb of RAM, 8 cores of type Intel(R) Xeon(R) CPU E5-2637 v3 @ 3.50GHz. RAID-Z3 with 12x HDD 10TB SAS 12Gb/s 7200 RPM. While we run hdt2rdf and hdtCat on this configuration this was not the case for HDT-MR. For the results of HDT-MR we report the ones achieved by \cite{gimenez-garcia_hdt-mr:_2015}, that where executed on a cluster with a total memory of 128Gb of RAM. While rdf2hdt and HDTCat are designed to be used on a single server, HDT-MR is designed to be used on a cluster. To make the results comparable we choose a single node and a cluster configuration with the same amount of RAM since this is the limited resource for compressing RDF serializations to HDT.
    \item \textbf{Configuration 2}: A server with 32 Gb of RAM, 16 cores of type Intel(R) Xeon(R) CPU E5-2680 0 @ 2.70GHz. RAID-Z3 with 12x HDD 10TB SAS 12Gb/s 7200 RPM.
    \item \textbf{Configuration 3}: A desktop computer with 16 Gb of RAM, AMD A8-5600K with 4 cores. 1x HDD  500GB SCSI 6 Gb/s, 7200 RPM.
\end{itemize}

Note that while the two first configurations have a RAID deployment with 10 drives, the third one is limited to a single HDD. Since HDTCat is I/O intensive, this can affect its performance.

The results obtained by the 3 methods on the 3 hardware configurations are compared in Table \ref{table-result}. It summarizes the comparison between the three methods to generate HDT from other RDF serializations on LUBM. \textbf{T} indicates the time and \textbf{M} the maximal memory consumption of the process. In the case of HDTCat we also report \textbf{$T_{com}$} the time to compress the N-Triples into HDT and \textbf{$T_{cat}$} the time to cat the two files together. $\star$ indicates that the experiment failed with an OUT OF MEMORY error. "-" indicates that the experiment was not performed. This has two reasons. Either a smaller experiment failed with an OUT OF MEMORY, or the experiment with HDT-MR was not performed on the corresponding configuration. The experiments in the T\_com column are very similar because we compress similar amount of data. We report the average times of these experiments and indicated that with "$\ast$". 

The results for Configuration 1 show that while hdt2rdf fails to compress lubm-12000, by using HDTCat we are able to compress lubm-40000. This means that one can compress at least as much as the HDT-MR implementation. Note that lubm-40000 does not represent an upper bound for both methods. For lubm-8000, HDT-MR is 121\% faster then HDTCat. This is expected since HDT-MR exploits parallelism while HDTCat does not. Moreover while the single node configuration has HDD disks, the cluster configuration used SSD disks. For lubm-40000 the speed advantage reduces, HDT-MR is 52\% faster then HDTCat.
The results for Configuration 2 show that the speed of hdtCat to compress lubm-40000 in comparison to Configuration 1 is reduced, but only by 25\%.
The results for Configuration 3 show that it is possible to compress on a 16Gb machine HDT files containing 5 Billion triples. In particular this means that it is possible to index on a 16Gb machine an RDF file with 5 Billion triples and construct a SPARQL endpoint on top. This is unfeasible for every other SPARQL endpoint implementation we are aware of. Moreover this also shows that for Configuration 1, lubm-4000 is far from being an upper bound so that potentially huge RDF files can be indexed, which was not imaginable before.\\

\textbf{Experiment 2} While the above results are using the syntetic data provided by LUBM we also performed an experiment using real datasets. In particular we join the Wikidata dump of the 19-02-2018 (330G in ntriple format) and the 2016 DBpedia dump\footnote{All files retrived by: wget -r -nc -nH --cut-dirs=1 -np -l1 -A '*ttl.bz2'  -A '*.owl'-R '*unredirected*'--tries 2 http://downloads.dbpedia.org/2016-10/core-i18n/en/, i.e. all files published in the english DBpedia. We exclude the following files: nif\_page\_structure\_en.ttl, raw\_tables\_en.ttl and page\_links\_en.ttl since they do not contain typical data used in application relying on DBpedia} (169G in ntriple format). This corresponds to 3.5 billion triples. We where able to join the corresponding HDT file in 143 minutes and 36s using a 32 Gb RAM machine. The maximal memory consumption was 27.05 Gb.\\

\textbf{Experiment 3} Note that Wikidata and DBpedia are not sharing many IRIs. So one valid argument is if HDTCat is also performing well when the two joined HDT files contain many common IRIs. To test this we randomly sorted the lubm.2.000.nt file and split it in two files containing the same amount of triples. We then join them using HDTCat. While joining lubm.1-1000.hdt and lubm.1001-2000.hdt took 287 seconds, joining the randomly sorted files took 431 seconds. This corresponds to a 66\% increase of time which is expected. This shows that HDTCat is still performing well in such a scenario.\\

\textbf{Code} The code is currently part of the HDT code repository available under \url{ https://github.com/rdfhdt/hdt-java}. The code is released under the \textit{Lesser General Public License} as the existing Java code.

\section{Conclusion and Future Work}
We have presented HDTCat, an algorithm and command line tool to join two HDT files which is time and memory efficient. We have described in detailed how the algorithm works and we have validated our implementation against the other two available alternatives namely rdf2hdt and HDT-MR.\\
In future we want to create a command line tool that combines rdf2hdt and HDTCat. rdf2hdt generates HDT files as a single threaded process. By combining it with HDTCat it will be possible to parallelize this and generate HDT files faster.\\
Moreover we would like to extend HDTCat to be able to merge more then two HDT files at the same time.\\
Another possible application is the use of HDTCat for constructing HDT triple-stores that support updates. Currently triple-stores that rely on HDT are read-only. A strategy used in modern databases is to have a read-only index and to store the updates in a delta structure that is periodically merged with the read only part.\\
Finally we believe that HDTCat will enable the Semantic Web Community to tackle scenarios which were infeasable before because of scalability. 

\paragraph{Acknowledgements:} We would like to thank Pedro Migliatti for executing part of the experiments as well as Javier D. Fernández for the helpful discussions with him. This project has received funding from the European Union's Horizon 2020 research and innovation program under the Marie Sklodowska-Curie grant agreement No 642795.


\bibliographystyle{splncs03}
\bibliography{zotero}

\end{document}